\documentclass[%
 preprint, 
 amsmath,amssymb,
 aps, physrev,
]{revtex4-2}

\usepackage{graphicx}
\usepackage{dcolumn}
\usepackage{bm}
\usepackage{hyperref}

\usepackage{multirow}%
\usepackage{amsmath,amssymb,amsfonts}%

\begin{document}

\title{\textbf{Topological Origin of Horizon Temperature via the Chern-Gauss-Bonnet Theorem} 
}%

\author{Jack C. M. Hughes}
\author{Fedor V. Kusmartsev}%
 \email{Contact author: fedor.kusmartsev@ku.ac.ae}
\affiliation{%
 College of Art and Science, Khalifa University, Abu Dhabi, PO Box 127788, UAE.
}%

\date{\today}

\begin{abstract}
This paper establishes a connection between the Hawking temperature of spacetime horizons and global topological invariants, specifically the Euler characteristic of Wick-rotated Euclidean spacetimes. This is demonstrated for both de Sitter and Schwarzschild, where the compactification of the near-horizon geometry allows for a direct application of the Chern-Gauss-Bonnet theorem. For de Sitter, a simple argument connects the Gibbon-Hawking temperature to the global thermal de Sitter temperature of the Bunch-Davies state. This establishes that spacetime thermodynamics are a consequence of the geometrical structure of spacetime itself, therefore suggesting a deep connection between global topology and semi-classical analysis. 
\end{abstract}

\maketitle

\section{Introduction}
It has been shown in many calculations that the laws of black hole mechanics \cite{Bardeen1973} are in fact thermodynamic relations, with the explanation for the entropy being of major interest \cite{Bekenstein1973, Hawking1975, Parikh2000, Akhmedov2006, Akhmedov2007, Wald1993, Ashtekar1998, Domagala2004, Bombelli1986, Srednicki1993, Strominger1996, Susskind1993, Susskind1994, Horowitz1993, Strominger1998, Carlip2014, Grumiller2023, MunozdeNova2019}. In general, spacetimes possessing a Killing horizon are thermodynamic bodies: both cosmic horizons (generated by $\Lambda$) and horizons associated to accelerated observers radiate with a thermal spectrum \cite{Unruh1976, GibbonsHawking1977, Gibbons1977}. For de Sitter however the situation is more complicated as a consequence of the horizon being observer dependent and the instability of the quantum state \cite{Ford:1984hs, Bros2008, Volovik2009, Polyakov2008b, Krotov2011, Jatkar2011, Polyakov2008, AndersonMottola2013, Dvali2018, Volovik2023, Diakonov2025, Volovik2025}. For black holes formed from gravitational collapse this lead Hawking to the information loss paradox \cite{Hawking1975, Hawking1974, Hawking1976}, which has been thoroughly investigated \cite{Mathur2009, Almheiri2013, Harlow2016, Chakraborty2017, Perez2017, Raju2021}. The recent replica trick provided crucial insight into the nature of the problem in the semi-classical regime \cite{Almheiri2020, Almheiri2020_2, Penington2022}.   

For both de Sitter and Schwarzschild the foundational derivation for the Hawking temperature is through the construction and analysis of the partition function for the gravitational path integral \cite{GibbonsHawking1977, Gibbons1977, Birrell1982, Padmanabhan2010, Ortin2004}. The basic idea is that we isolate a static patch of the spacetime and perform a Wick rotation to Euclidean time (analytic continuation $t \to -i\tau$). Then, the near horizon limit will be smooth only if $\tau$ is periodic. Viewed as a thermal field theory \cite{Birrell1982, Kubo1957, MartinSchwinger1959}, this periodicity maps to the temperature of radiation. 

A number of attempts have been made to develop a topological basis to the Hawking temperature (and thermodynamics more generally), following the observations of the extreemal cases \cite{Hawking1995, Gibbons1995, Padmanabhan2002, Volovik2025}. For the three-dimensional BTZ black hole, this approach is natural in the context of Witten's reformulation\cite{Witten1988, Banados1994, Zhang2023}. The emphasis is placed on the Chern-Gauss-Bonnet theorem due to its relationship with boundary contributions \cite{Gibbons1995, Banados1994, Liberati1997, Robson2019, Ovgun2020, AltasTekin2023} (see also \cite{Liberati1998, Xian2023}) and Lovelock gravity \cite{Padmanabhan2010, Banados1994, Lovelock1972, Fernandes2022, Padmanabhan2013}. 

These results are often specialized - since the theorem applies only to compact manifolds \cite{Chern1944, Chern1945, Mielke2017, Besse1987}, hence the typical need for a boundary - with unclear physical basis in relation to the Gibbons-Hawking construction and the Wick rotation $t \to -i\tau$. The purpose of this work is to demonstrate explicitly that for both Schwarzschild \textit{and} de Sitter, the Hawking temperature is of topological origin. The Wick rotation and resulting periodicity of Euclidean time makes the manifold compact in the near-horizon limit, allowing the theorem to be applied without the need for boundaries. In other words, the Hawking temperature $T_H$ maps to the Euler characteristic $\chi(\mathcal{M})$ directly as a consequence of $t \to -i\tau$ and periodicity in $\tau$. This result therefore provides a unified topological framework for the Hawking temperature in the non-extreemal case. 

\section{Background}
Any spacetime $\mathcal{M}$ satisfying the Einstein field equations in vacuum 
\begin{equation}
    R_{\mu \nu} - \frac{1}{2} R g_{\mu \nu} + \Lambda g_{\mu \nu} = 0
\end{equation}
is required to have Ricci tensor proportional to the metric
\begin{equation}\label{EC}
    R_{\mu \nu} = \Lambda g_{\mu \nu}. 
\end{equation}
By definition, such spacetimes are Einstein manifolds \cite{Besse1987}. This has important consequences for the geometry of $\mathcal{M}$ \cite{Atiyah1978, Krasnov2011}. The Riemann tensor $R_{\mu \nu \rho \sigma}$ admits its orthogonal Ricci decomposition into its Weyl $C_{\mu \nu \rho \sigma}$, Ricci $R_{\mu \nu}$ and Ricci scalar $R$ components \cite{Weinberg1972}:
\begin{equation}
    \begin{split}
        R_{\mu \nu \rho \sigma} = &\;  C_{\mu \nu \rho \sigma} + \frac{1}{(n-2)} \big( g_{\mu \rho} R_{\nu \sigma} - g_{\mu \sigma} R_{\nu \rho} + g_{\nu \sigma} R_{\mu \rho} - g_{\nu \rho} R_{\mu \sigma}\big) \\
        & + \frac{R}{(n-1)(n-2)}\big(g_{\mu \rho} g_{\nu \sigma} - g_{\mu \sigma} g_{\nu \rho}\big)
    \end{split} 
\end{equation}
where $n$ is the dimension of spacetime. With (\ref{EC}) satisfied, the Riemann tensor is now expressed as (setting $n = 4$)
\begin{equation}\label{Rie_Decomp}
    R_{\mu \nu \rho \sigma} = C_{\mu \nu \rho \sigma} +  \frac{\Lambda}{3}\big(g_{\mu \rho} g_{\nu \sigma} - g_{\mu \sigma} g_{\nu \rho}\big).
\end{equation}
Therefore, once the metric symmetry is fixed the remaining freedom for the curvature is in the algebraic decomposition of the Weyl tensor (through the Petrov classification) \cite{Petrov1969, Petrov2000, Pirani1957}. 

In four dimensions (and more generally any even dimension spacetime), the curvature of a vacuum solution is additionally constrained by the topology of spacetime. This is a consequence of the Chern-Gauss-Bonnet theorem, which relates the integral of the curvature 2-form $\mathcal{R}$ to the topological Euler characteristic $\chi(\mathcal{M})$ of spacetime \cite{Chern1944, Chern1945, Mielke2017}. In tensor indices, this is expressible as 
\begin{equation}
    \chi(\mathcal{M}) = \frac{1}{32\pi^2} \int_{\mathcal{M}} \sqrt{-g} d^4x  \big(R^2 - 4R_{\mu \nu}R^{\mu \nu} + R_{\mu \nu \rho \sigma}R^{\mu \nu \rho \sigma}\big).
\end{equation}
This term coincides with the Gauss-Bonnet Lagrangian density known from Lovelock gravity \cite{Padmanabhan2013, Lovelock1972, Fernandes2022}. Its variation vanishes identically in four dimensions and hence does not affect the Einstein equations if included in the action principle \cite{Corichi2016}, while in higher dimensions it need not vanish.  For a vacuum spacetime (\ref{EC}), the first two terms cancel, leaving
\begin{equation}\label{CGB_Rie}
    \chi(\mathcal{M}) = \frac{1}{32\pi^2} \int_{\mathcal{M}} \sqrt{-g} d^4x  \; R_{\mu \nu \rho \sigma}R^{\mu \nu \rho \sigma}.
\end{equation}
Expressing everything instead in terms of the decomposition (\ref{Rie_Decomp}), this is
\begin{equation}
    \chi(\mathcal{M}) = \frac{1}{8\pi^2} \int_{\mathcal{M}} \sqrt{-g} d^4x \bigg(\frac{1}{4} C_{\mu \nu \rho \sigma}C^{\mu \nu \rho \sigma} +  \frac{2}{3} \Lambda^2 \bigg).
\end{equation}

The difficulty in the application of this theorem is that it requires $\mathcal{M}$ to be \textit{compact} (closed and bounded) \cite{Chern1944, Chern1945}. For most vacuum spacetimes of interest (e.g. Schwarzschild and de Sitter) this is not the case. The theorem will still constrain the curvature spectrum $R_{\mu \nu \rho \sigma}R^{\mu \nu \rho \sigma}$ in any locally compact domain, but globally the integral (\ref{CGB_Rie}) will not be restricted to the topological integer $\chi(\mathcal{M})$. Various tricks allow this to be mapped to submanifold contributions and to allow for boundaries \cite{Chern1945}, which is what has been explored in the previous calculations \cite{Banados1994, Liberati1997, Robson2019, Liberati1998}. In two dimensions the theorem is applied after a Wick rotation, with Euclidean time compactified and such a submanifold identified through the time-like Killing vector \cite{Zhang2023, Robson2019, Xian2023}. 

The idea of this paper is to demonstrate that for de Sitter and Schwarzschild in four dimensions such additional arguments are unnecessary. The theorem may be directly applied in the Wick rotated Euclidean space where compactness is \textit{enforced} through the appropriate near-horizon limit. The Hawking temperature is set by the Euler characteristic $\chi(\mathcal{M})$, and hence is of topological origin. For de Sitter, this gives a clear geometrical origin for the difference between the Gibbons-Hawking temperature $T_H = 1/2\pi l$ and the bulk de Sitter temperature $T_H = 1/\pi l$ \cite{Volovik2023}.

\section{de Sitter}
Let us take the de Sitter spacetime in static coordinates \cite{Hawking1973, Gron2007}
\begin{equation}\label{dS}
    ds^2 = -\bigg(1 - \frac{r^2}{l^2}\bigg) dt^2 + \bigg(1 - \frac{r^2}{l^2}\bigg)^{-1}dr^2 + r^2 d\Omega_{\mathbb{S}^2},
\end{equation}
where $l$ is expressed in terms of the cosmological constant $\Lambda$ via $l^2 = 3/\Lambda$. Note that the cosmological horizon is the hypersurface swept by $r = l$. We wish to evaluate the integral 
\begin{equation}\label{I_G}
    I_G = \frac{1}{32\pi^2} \int \sqrt{-g} d^4 x \; R_{\mu \nu \rho \sigma} R^{\mu \nu \rho \sigma},
\end{equation}
which is not restricted to $\chi(\mathcal{M})$ since de Sitter is not compact. However, let us Wick rotate to Euclidean time $t \to i \tau$; the metric takes the form 
\begin{equation}\label{Euclidean_dS}
    ds_E^2 = \bigg(1 - \frac{r^2}{l^2}\bigg) d\tau^2 + \bigg(1 - \frac{r^2}{l^2}\bigg)^{-1}dr^2 + r^2d\Omega_{\mathbb{S}^2}.
\end{equation}
The horizon is no longer a null hypersurface in the sense of (\ref{dS}). Instead, $r = l$ is a non-regular region of the geometry that must be `smoothed out' to eliminate the conical singularity. To see this, we introduce a new radial expansion relevant for the near-horizon reparameterization:
\begin{equation}
    r = l(1 - \rho^2).
\end{equation}
The metric function therefore expands as 
\begin{equation}
    1- \frac{r^2}{l^2} \approx2\rho^2
\end{equation}
and to leading order, the near-horizon limit of (\ref{Euclidean_dS}) is
\begin{equation}\label{NHdS}
     ds_E^2 = 2l^2\bigg(\frac{\rho^2}{l^2} d\tau^2 + d\rho^2\bigg) + l^2 d\Omega_{\mathbb{S}^2}.
\end{equation}
The metric is now of the standard Rindler form and admits a conical singularity as $\rho \to 0$. This conical singularity can be eliminated provided we identify $\tau \sim \tau + \beta$ (i.e. compactify the Euclidean time) with period
\begin{equation}
    \beta = 2 \pi l.
\end{equation}
This is the standard Gibbons-Hawking argument to construct the temperature $T_H = 1/ \beta$ \cite{GibbonsHawking1977, Gibbons1977}.

The $(\tau, \rho)$ plane becomes topologically the disc $D^2$, which with the spherical line element $d\Omega_{\mathbb{S}^2}$ indicates the near horizon geometry (\ref{NHdS}) is compact $D^2 \times \mathbb{S}^2$. That is, (\ref{I_G}) can now be explicitly computed to yield the Euler characteristic of the Wick rotated geometry, 
\begin{equation}
    I_G = \chi(\mathcal{M}_E) = \frac{1}{32\pi^2} \int_{\mathcal{M}_E} \sqrt{-g} d^4 x \; R_{\mu \nu \rho \sigma} R^{\mu \nu \rho \sigma}.
\end{equation}
The coordinate system in which we choose to compute the integral is irrelevant provided we respect the compact domain $(\tau, \rho)$ enforced in the near-horizon limit. That is, the Kretschmann scalar $R_{\mu \nu \rho \sigma} R^{\mu \nu \rho \sigma}$ is a curvature invariant \cite{Hervik2010}. This allows us to write 
\begin{equation}\label{GBT_dS_Half}
    \chi(\mathcal{M}_E) =  \frac{1}{32\pi^2} \int^{\beta}_0 d\tau \int^{l}_{0} dr \;  \int_{\mathbb{S}^2} \sqrt{g} d^2 x \; R_{\mu \nu \rho \sigma} R^{\mu \nu \rho \sigma}.
\end{equation}
Identifying $\beta = 1/T_H$, we choose to represent this as a formula for the Hawking temperature:
\begin{equation}
    T_H = \frac{1}{32\pi^2 \chi(\mathcal{M}_E)}  \int^{l}_{0} dr \;  \int_{\mathbb{S}^2} \sqrt{g} d^2 x \; R_{\mu \nu \rho \sigma} R^{\mu \nu \rho \sigma}.
\end{equation}
For the de Sitter geometry,
\begin{equation}
    R_{\mu \nu \rho \sigma} R^{\mu \nu \rho \sigma} = \frac{24}{l^4},
\end{equation}
and the above evaluates to
\begin{equation}
    T_H = \frac{1}{\chi(\mathcal{M}_E) \pi l}.
\end{equation}
Finally the Euler characteristic of a topological product is the product of Euler characteristics of the individual spaces
\begin{equation}\label{Euler_Disc}
    \chi(\mathcal{M}_E) = \chi(D^2 \times \mathbb{S}^2) = \chi(D^2) \cdot \chi(\mathbb{S}^2) = 1\cdot 2 = 2,
\end{equation}
which reproduces the Gibbons-Hawking temperature of de Sitter via the topological Chern-Gauss-Bonnet theorem.

There is a clear geometrical argument within this picture of the factor of $2$ that separates the Gibbons-Hawking temperature $T_H = 1/2 \pi l$ and the thermal de Sitter temperature $T_H = 1/\pi l$ (the latter of these is motivated through the periodicity of Wightman functions for the Bunch-Davies vacuum \cite{Birrell1982, Diakonov2025, BunchDavies1978, Akhmedov2020}). The computation (\ref{GBT_dS_Half}) is for the near horizon limit of the Euclidean de Sitter spacetime. This is the static patch representation of Wick rotating global de Sitter spacetime to the Euclidean 4-sphere. Indeed, globally Euclidean de Sitter has topology $\mathbb{S}^4$, but the coordinates $(\tau, \rho)$ cover only \textit{half} this space in the static patch with topology $D^2 \times \mathbb{S}^2$. Concretely, the smooth point $\rho = 0$ represents the equatorial line of $\mathbb{S}^4$ viewed as the continuous matching of two such hemispheres. As a consequence if we want to compute the Euler characteristic of the \textit{total} Euclidean de Sitter topology, the integral contribution (\ref{GBT_dS_Half}) must be doubled. Topologically this is clear \cite{Hatcher2002}, and at the physical level de Sitter being maximally symmetric ensures both hemispheres contribute equally to $\chi({\mathbb{S}^4})$ via the homogeneity of curvature. Computationally then, we have
\begin{equation}
    \chi({\mathbb{S}^4}) =  \frac{1}{16\pi^2 T_H} \int^{l}_{0} dr \;  \int_{\mathbb{S}^2} \sqrt{g} d^2 x \; R_{\mu \nu \rho \sigma} R^{\mu \nu \rho \sigma} = \frac{2}{\pi l T_H}.
\end{equation}
But noting that $ \chi({\mathbb{S}^4}) = 2$, we recover the thermal de Sitter temperature 
\begin{equation}
    T_H = \frac{1}{\pi l}
\end{equation}
through the topological Chern-Gauss-Bonnet theorem.

\section{Schwarzschild}
The case for Schwarzschild proceeds in much the same way. The main difference is that the compactification of $\tau$ `bounds' the radial coordinate from the horizon \textit{outwards}, meaning that there is infinite range radially. The Wick rotated line element is 
\begin{equation}\label{E_S}
    ds_E^2 = \bigg(1 - \frac{2M}{r}\bigg) d\tau^2 + \bigg(1 - \frac{2M}{r}\bigg)^{-1} dr^2 + r^2 d\Omega_{\mathbb{S}^2},
\end{equation}
which again possesses a conical defect that must be smoothed in the near-horizon limit. We can expand linearly 
\begin{equation}
    r = 2M + \rho
\end{equation}
so that the metric function approximates to 
\begin{equation}
    1 - \frac{2M}{r} \approx \frac{\rho}{2M}.
\end{equation}
The Rindler form for (\ref{E_S}) is identified via setting $\rho =R^2/ {8M}$, such that
\begin{equation}
    ds_E^2 \approx \frac{R^2}{16M^2} d\tau^2 + dR^2 + 4M^2 d\Omega_{\mathbb{S}^2}.
\end{equation}
In accordance with the Gibbon-Hawking method \cite{GibbonsHawking1977, Gibbons1977}, the temperature is found by demanding periodicity in $\tau$:
\begin{equation}
    \tau \sim \tau + \beta, \quad \beta = 8\pi M,
\end{equation}
which ensures the point $r = 2M$ is regular in the Euclidean geometry. 

Again, this is the disc topology in $(\tau, r)$ allowing us to implement the Gauss-Bonnet theorem in the near-horizon limit. Now, Euclidean Schwarzschild is clearly non-compact since $r \in [2M, \infty)$. However this point is irrelevant in the computation of the Euler characteristic since the Kretschmann scalar $R_{\mu \nu \rho \sigma} R^{\mu \nu \rho \sigma}\sim 1/r^6$ falls of sufficiently fast for the integral to be \textit{asymptotically compact}. In other words, the coordinate system respects the near-horizon topology: there is no contribution from the asymptotic element to $\chi(\mathcal{M}_E)$. Explicitly, 
\begin{equation}
  \chi(\mathcal{M}_E) =  \frac{1}{32\pi^2} \int^{\beta}_0 d\tau \int^{\infty}_{2M} dr \;  \int_{\Sigma} \sqrt{g} d^2 x \; R_{\mu \nu \rho \sigma} R^{\mu \nu \rho \sigma}.
\end{equation}
Recalling
\begin{equation}
    R_{\mu \nu \rho \sigma} R^{\mu \nu \rho \sigma} = \frac{48 M^2}{r^6},
\end{equation}
we observe 
\begin{equation}
    T_H = \frac{1}{\chi(\mathcal{M}_E)4\pi M}.
\end{equation}
Following again the logic of (\ref{Euler_Disc}), this verifies the Hawking temperature and Euler characteristic are dual to eachother. 

\section{Conclusion}
This work demonstrates that there exists a unified topological principle underlying the thermodynamics of Killing horizons in four-dimensional vacuum spacetimes. The Hawking temperature is inversely proportional to the Euler characteristic of the Wick rotated Euclidean geometry in its smooth compact near horizon limit. This is made possible through the application of the Chern-Gauss-Bonnet theorem \cite{Chern1944, Chern1945, Mielke2017} without the need for submanifold or boundary arguments. The de Sitter case is particularly interesting, since the calculation of the Gibbons-Hawking temperature $T_H = 1/2\pi l$ can be immediately extended to the (global) thermal de Sitter temperature $T_H = 1/\pi l$ through a simple topological argument. Indeed, the near-horizon topology $D^2 \times \mathbb{S}^2$ in (\ref{GBT_dS_Half}) covers half of the 4-sphere (global Euclidean de Sitter), and so a calculation respecting $\chi(\mathbb{S}^4)$ requires \textit{double} the Hawking temperature for consistency. This is likely the origin of observed bulk-boundary correspondence in the de Sitter entropy \cite{Volovik2023, Diakonov2025, Volovik2025}. Extending these results to Kerr is a non-trivial problem: the Wick rotation must be supplemented by an appropriate redefinition of the azimuthal coordinate to ensure off-diagonal elements of the metric remain real. However, the methodology presented here is likely an underlying topological argument supporting the results of Altas \& Tekin \cite{AltasTekin2023}.

\bibliographystyle{apsrev4-2}
\providecommand{\noopsort}[1]{}\providecommand{\singleletter}[1]{#1}%
%

\end{document}